# Fully atomic layer deposited transparent carrier-selective contacts for bifacial Cd-free Cu$_2$ZnSnSe$_4$ thin-film solar cells

*Rosa Almache-Hernández[a]*[‡], Gerard Masmitjà[b]*[‡], Benjamín Pusay[a], Eloi Ros[b], Kunal J. Tiwari[b], Pedro Vidal-Fuentes[c], Victor Izquierdo-Roca[c], Edgardo Saucedo[b], Cristóbal Voz[b], Joaquim Puigdollers[b] and Pablo Ortega[b]*

[a] School of Physical Sciences and Nanotechnology, Yachay Tech University, Urcuquí 100119, Ecuador

[b] Electronic Engineering Department, Universitat Politècnica de Catalunya, Jordi Girona 1 – 3, 08034, Barcelona, Spain

[c] Catalonian Institute for Energy Research (IREC), Barcelona, Spain

ABSTRACT

Thin-film solar cells based on kesterite ($Cu_2ZnSnSe_4$) material are a promising alternative for photovoltaic devices due to their composition consisting of earth abundant elements, ease of production at a relatively low temperatures and excellent optical absorption properties. Additionally, this absorber compound allows a tuneable bandgap energy in the 1 to 1.5 eV window range, which makes it an attractive candidate either as a top or a bottom solar cell in tandem technologies combined with transparent carrier-selective contacts. However, conventional kesterite devices use a toxic CdS layer as an electron-selective contact, resulting in the difficult-to-dispose chemical waste. This work explores the use of a stack of ZnO and Al-doped ZnO (AZO) films deposited by ALD to replace the CdS-based contacts in kesterite devices. The inclusion of a polyethylenimine (PEI) interlayer as dipole to enhance the overall electrical contact performance is also discussed. The transparent back contact is formed by an ALD $V_2O_x$ thin layer over a FTO conductive electrode. Fabricated kesterite solar cells exhibit remarkable photocurrent density values of 35 $mAcm^{-2}$, open-circuit voltage around 260 mV and efficiencies up to 3.5% using front illumination. The aforementioned photovoltaic parameters yield to 5.3 $mAcm^{-2}$, 160 mV and 0.3% respectively under back illumination, demonstrating the bifaciality of the proposed structure.

INTRODUCTION

The global energy demand has been on a continuous growth since the industrial revolution by the end of the 18[th] century. In the upcoming years, it will be necessary to shift from traditional energy sources towards more sustainable and environmentally friendly alternatives. In this regard, the adoption of renewable energy sources, such as photovoltaic energy, has emerged as a significant imperative for both industries and households worldwide [1], [2], [3]. As a clean and renewable energy source, photovoltaic energy provides an ideal solution to meet the growing energy needs while minimizing the environmental impact [4].

While silicon based photovoltaic technology has been the dominant mainstream technology in the industry [5], [6], nowadays, there is an increase in interest in thin-film solar cells. These devices are composed of absorbers which are only ~2 μm thick, a remarkable 100 times thinner than traditional silicon-based substrates [7], [8]. The potential benefits of thin-film solar cells are becoming increasingly attractive, including their lower cost, lighter weight, and flexibility. As a result, a significant research and development effort is underway to improve the efficiency and scalability of thin-film solar cells for widespread adoption in the energy industry [9], opening the door to fabricate cheaper, lighter and more flexible solar cells.

Kesterite, also known as $Cu_2ZnSn(S,Se)_4$, is a material that shows excellent properties as an absorber layer for solar cells [10]. Its high absorption coefficient and ease of synthesis makes it particularly attractive for photovoltaic applications [11], [12]. The high light absorption of this material is due to its crystal structure similar to chalcopyrite [13], [14], which enables it to efficiently capture a broad range of light wavelengths in the UV-Vis-NIR part of the solar spectrum. Additionally, it can be synthesized using low-cost and low-temperature methods, which makes it an economical and environmentally friendly alternative to other thin-film materials such as cadmium telluride (CdTe) or copper indium diselenide (CIS) compounds [4], [7], [15]. As a result, there is considerable interest in researching and developing kesterite-based solar cells as a promising new technology in the renewable energy industry [16]. Additionally, this material has, among other properties, a tunable and direct bandgap (1 to 1.5 eV) and critical raw materials (CRM)–free composition [17], [18]. Band gap tuning depends on the presence of selenium, sulfur or both within its chemical composition [19]. Co-sputtering is currently the most widely applied method for synthesizing kesterite absorbers along with the emerging solution based chemical routes. It is considered as the most feasible and practical approach to be applied in the PV industry [20]. However, other methods for kesterite synthesis have also been reported, including electrodeposition of metal precursors [21], co-evaporation [22] and spin-coating techniques [23]. While these methods have shown some promise in the laboratory, they still require further optimization to achieve the efficiency, scalability and cost-effectiveness necessary for a large-scale production.

Despite the high potential of kesterite as PV material, the efficiency record achieved by this technology stands at a moderate 14.9% value compared to other technologies [24]. One of the major obstacles to improve the efficiency of kesterite solar cells is the open circuit voltage ($V_{oc}$) losses, which is often attributed to interface recombination at the contacts of the structure [25]. To address this challenge, researchers are focusing on selecting the appropriate materials for both the electron transport layer (ETL) and the hole transport layer (HTL) applied to kesterite solar cells. By optimizing these contacts, it may be possible to reduce interface recombination and improve the overall efficiency.

In this regard, transition metal oxides (TMOs) have been successfully utilized as carrier-selective contacts in different crystalline silicon solar cell technologies [26], [27]. These materials are semiconductors with high bandgap energies (~3 eV) offering a high transparency as well as an excellent carrier-selective property to block one type of carrier (electrons or holes). The use of TMOs as carrier-selective contacts represents an exciting development in the field of solar cell technology. The continued research in this area holds significant promise to achieve higher efficiency and lower costs in the solar cell production. Recently, TMOs have been applied as selective contacts in kesterite solar cells and encouraging results have been reported [28], [29],

[30]. This approach has demonstrated its potential to improve the efficiency of kesterite solar cells by enabling efficient carrier extraction.

Cadmium sulfide (CdS) is the standard ETL material applied to kesterite solar cells. CdS compound has a bandgap of approximately 2.4 eV [31], being the typical thickness in the 30-70 nm range [32]. Therefore, residual optical absorption in the CdS layer allows only about 80% light transmittance to the absorber [33]. Moreover, unfavorable energy band alignment at the CdS/CZTSe interface [34], [35] causes additional recombination losses and limits the fill factor. Another important drawback of CdS films is the well-known toxicity of Cd [36] and the toxic waste generated during their deposition stage, which makes the CdS compound an unsuitable material for widespread use in the PV industry. Hence, it is necessary to develop alternative ETL materials with high transmittance, favorable energy band alignment with the absorber and low toxicity to reduce environmental impact. With this idea in mind, ETLs based on TMOs seem to be also a natural choice to replace CdS-based selective contact [37]. In perovskite and organic photovoltaics, the application of ZnO as electron transport layer suggests a high performance of this material in the selection and transport of electrons during solar cell operation [38], [39]. Several studies have already reported the replacement of CdS as an ETL in the CZTSe and CIGS solar cells by ZnO films deposited by either atomic layer deposition (ALD) or sputtering techniques in combination with aluminum-doped zinc oxide (AZO) electrodes, achieving a good solar cell performance [40], [41].

Recently, some research studies have shown that the intercalation of very thin-film layers exhibiting dipole moment can modify the work function of the metal and improve the band alignment at the metal/semiconductor junction [42], [43]. There are several compounds that have intrinsic dipole moments and are compatible with device manufacturing. In most of them (e.g.,

PEI, PFN, PEIE) the dipole moment is due to the presence of amino groups (-NH$_2$). In this work we have focused on the use of polyethylenimine (PEI) amino acid polymer, due to its good results reported elsewhere with crystalline silicon absorbers. In ref. [44], the use of PEI as polymeric selective contact increases the efficiency of c-Si based photovoltaics devices from 14.2% to 16%. Other studies reported the use of an interlayer dipole to enhance contact selectivity in other solar cell technologies, either individually [45] or together with TMOs layers [46].

The standard HTL on kesterite solar cells is based on a MoSe$_2$ layer on the back of the device. The MoSe$_2$ film grows spontaneously during the selenization stage during the kesterite synthesis process [47], [48] . However, this HTL has the main drawback to be opaque to the light. Developing transparent selective contacts for thin-film solar cells has become an important challenge to fabricate semi-transparent [49], bifacial solar cells [13] or as contact for the top solar cell in tandem structures [50]. In the previous studies, it was reported the possibility to grow CZTSe onto semi-transparent Mo (12 nm) layers [51] and even directly onto FTO-based conductive glass as an alternative to the Mo metal electrode [52]. The FTO material stands out, among other transparent conductive oxides, for its robustness against high temperature processes. Although the optical transmittance is known to decrease from 80% to 76% at 550 nm due to the kesterite selenization process [53], the FTO sheet resistance appears to be perfectly stable at high temperatures (up to 600 ºC) [54], offering greater efficiency and stability than the Mo-based counterparts after a standard damp heat test [55]. Replacing Mo with FTO conductive electrodes calls for the need of alternative HTL materials. In this work, kesterite was synthesized onto Glass/FTO/V$_2$O$_x$ layers, where the V$_2$O$_x$ film, deposited by ALD technique acts as the transparent hole-selective contact. V$_2$O$_x$ films have been successfully used as HTL on c-Si [56] solar cells and other thin-film technologies [49], [57]. One advantage of incorporating a transparent HTL in the back of thin-film solar cells is the possibility to absorb light from the albedo radiation (bifacial solar cell). Bifacial devices in combination with transparent or semi-transparent absorbers are very interesting for building integrated photovoltaic technologies (BIPV) [58] and agrivoltaic systems [59], allowing to integrate solar cells to the façades and/or windows of buildings and also in indoor photovoltaic applications due to their bifacial absorption ability [60].

In this study, we report on the fabrication of CZTSe photovoltaic devices based on TMOs as transparent carrier-selective contacts deposited by ALD. Concretely, the ETL layer is performed by ZnO/AZO stack and the HTL is carried out by a $V_2O_x$ layer. To improve the performance of electron-selective contact, the incorporation of a PEI-based dipole layer between the AZO-based transparent conductive oxide (TCO) electrode and the ZnO film is also evaluated. Device performance using conventional kesterite solar cells based on CdS and MoSe as ETL and HTL, respectively, is also compared.

EXPERIMENTAL SECTION

*A. CZTSe Photovoltaic Device Fabrication*: at first, commercial off-the-shelf FTO-coated glass substrates were cleaned using a sequence of water, acetone, and isopropanol baths. Then, $V_2O_x$ layer was deposited on top of the substrates using a thermal ALD system (Savannah S200, Cambridge Nanotech ALD system) with tetrakis(ethylmethylamino)-vanadium (IV) (VTIP from Sigma Aldrich with a >99.99% purity) and deionized water (DI-$H_2O$ with a resistivity of 16 MΩcm) as the vanadium precursor and oxidant species, respectively. The VTIP cylinder precursor and the process chamber were heated at 58 °C and 125 °C, respectively. The ALD process sequence of DI-H$_2$O/purge/VTIP/purge was set to 2/5/5/5 seconds and using N$_2$ as the carrier gas (20 sccm). A total of 200 ALD cycles were applied to the samples to obtain a final film thickness of about 10 nm.

Afterwards, to form the CZTSe absorber on the V$_2$O$_x$ layer, a metallic precursor stack was sputtered by a sequential deposition of Cu/Sn/Cu/Zn films with thicknesses of 20/290/190/175 nm and using a DC power of 100/50/100/100 W. The pressure inside the Alliance Concept AC-450 DC sputtering chamber was maintained at 3.3×10$^{-3}$ mbar. The optimized stack targeted compositions were of Cu/(Zn+Sn) = 0.72 and Zn/Sn = 1.05. Subsequently, the precursor stack was subjected to a two-stage reactive thermal annealing (RTA) treatment in the Hobersal three-zone tubular furnace. The samples were loaded into a graphite box with a 100 mg of Se and 5 mg of Sn, thus, maintaining the Se atmosphere during the whole thermal process as well as preventing the loss of Sn from the precursor stack. The first stage of the thermal treatment was performed at a temperature of 400 °C and a constant Ar flux during 30 min. The pressure inside the tube was maintained at 1.5 mbar. For the second stage, the temperature was raised to a value of 550 °C maintaining a constant pressure of 950 mbar for 15 min. Finally, the system was allowed to cool down naturally, that is 20 °C/min, and the samples were extracted at room temperature. The composition and thickness of the completed CZTSe absorber layer were determined through a pre-calibrated X-ray fluorescence (XRF) measurement system. The ratios of the atomic species were found to be Cu/(Zn+Sn) = 0.71 and Zn/Sn = 1.1 with a total thickness of the absorber about 1.4 µm.

Next, samples were cleaned with a 1% dilute hydrofluoric acid (HF) dip for 30 seconds to remove the secondary phases from the CZTSe surface [49].

For the ETL structure, firstly a thin layer of ZnO was deposited by the same ALD Savanna equipment, using diethylzinc (DEZ from UP Chemical Co. Ltda with a purity of 95%) as zinc precursor and DI-$H_2O$ as oxidizing agent. A total of 115 cycles were used to achieve a thickness of about 20 nm. ZnO films were deposited at different chamber temperatures, namely, 100, 110 and 125 ºC, setting the DEZ cylinder at room temperature in all cases. In some samples, a PEI layer with a 0.01% PEI/MetOH solution was spin-coated at 5000 rpm for 50 s over the ZnO layer. In the other devices, the dipole layer was deposited directly over the absorber surface without ZnO. Samples processed with dipole underwent a subsequent 5 min annealing at 80 ºC to promote the evaporation of the solvent, using a Selecta Rectangular precision hotplate Plactronic. Next, a transparent front contact based on aluminum doped zinc oxide (AZO), was deposited by ALD at 165 °C. Al doping of ZnO films was done using trimethyl aluminum (TMA from UP Chemical Co. Ltda with a purity of 99.99%) and using the super-cycle concept with a DEZ:TMA cycle ratio of 19:1. A total of 65 super cycles were deposited, reaching an AZO film thickness of ~200 nm with a sheet resistance in the range of 80 – 100 Ω/sq and n-type behavior (see Figure S1 in the Supporting Information). Finally, 150 nm thick silver fingers were thermally evaporated on top of the AZO film by means of a magnetic metal mask. Metal evaporation was performed using silver pellets (Sigma Aldrich; 99.99% purity) under high vacuum conditions ($10^{-5}$ mbar).

In all cases, isolated devices were defined with an active area of 0.3 cm x 0.3 cm by scribing the surface with a diamond tip.

Solar cell PV performance was also compared with a conventional kesterite solar cell (same area) fabricated using a Mo coated soda-lime glass substrate (SLG/Mo). In this case, the baseline process starts with deposition of a Mo layer (~800 nm) onto the SLG substrate by sputtering technique with a DC power of 330 W and a pressure inside the chamber of $1.5\times10^{-3}$ mbar. The absorber layer was grown in the same way as explained earlier. Note that a MoSe film in the Mo/CZTSe interface grows spontaneously during the selenization stage during the kesterite synthesis process. Next, the sample was cleaned in a bath of $KMnO_4 + H_2SO_4$ and $(NH_4)_2S$ with a subsequent KCN bath to remove impurities and secondary phases from the surface. After the deposition of the CdS layer (60 nm) by chemical bath deposition (CBD) technique, a stack of ZnO/ITO (50 nm/350 nm) was deposited at 200 °C by pulsed DC sputtering as a transparent top electrode.

**B. Device characterization:** photovoltaic *J-V* curves of the fabricated solar cells were measured in dark and under one-sun illumination (ORIEL 94021A Newport solar simulator) using standard test conditions (AM1.5G solar spectrum 1 $kWm^{-2}$ and 25 ºC). External quantum efficiency (EQE) was measured using a spectral response measurement system device (QEX10 PV Measurement system) with a voltage and white light bias of 0 V and 0.2 Suns, respectively.

**C. Material characterization and compositional analysis:** optical properties were studied using UV-VIS spectrometer Lambda 950 By PerkinElmer. High angle annular dark field (HAADF) scanning TEM images, in combination with energy dispersive X-ray spectroscopy (EDS) chemical composition analysis were executed by using a Cs-corrected FEI Titan (60 – 300 kV) transmission electron microscope. The FEI Titan was equipped with a high brightness Schottky emitter source (X-FEG gun), a monochromator and a Gatan 2k × 2k CCD camera. The cross-section lamellas were prepared from contact test samples, using a focused ion beam (FIB) lift-out technique. The samples were subjected to an oxygen plasma cleaning process before introducing them into the FEI Titan equipment.

XRF measurements were performed using Fischerscope model XDV instrument with X-ray source energy 50 KV equipped with Ni10 filter. The XRF system was pre-calibrated with ICP-OES (inductively coupled plasm optical emission spectroscopy) measurements. On each sample, the XRF data were collected at 5 different points to have a statistical estimation of the atomic percentage value of each constituent element of the CZTSe absorber.

RESULTS AND DISCUSSION

*A. ZnO-based ETL and dipole effect on solar cell performance*: To determine the performance of ALD ZnO layers as ETL in the CZTSe solar cells, several devices were fabricated using three different ETL configurations: devices with a single layer of ZnO (labelled S01), samples with only a PEI dipole deposited directly over the kesterite absorber (labelled S02) and devices with a stack of ZnO/PEI (labelled S03). The deposition temperature of ZnO films during the ALD stage was 110 °C, independently of the structure. In all cases, the HTL material was $V_2O_x$ deposited by ALD, as it is explained in the experimental section. As it can be seen in the sketch of the S03 structure in Figure 1a, the use of both transparent electrodes allows front and alternatively rear illumination, resulting in a bifacial solar cell structure.

Figure 1b shows the *J-V* characteristics under front illumination for the best-fabricated device in each S01, S02, and S03 structures. Table 1 summarizes the photovoltaics parameters of these devices, where $V_{oc}$, $J_{sc}$ *FF* and $\eta$ are the open circuit voltage, short-circuit current density, fill factor and efficiency, respectively.

Results demonstrate the relevant role of the ZnO layer to select and transport the electrons in the solar cell operation. Thus, the ALD ZnO film can be used not only as a buffer layer to avoid short circuits [61], but also as ETL due to its inherent n-type conductivity and wide band gap property [62], [63]. Devices using only the ZnO layer as ETL achieve efficiencies up to 1.3%, while devices with only a PEI dipole layer yield a poor efficiency of only 0.08%. However, the best approach is based on the ZnO/PEI stack (S03 devices), reaching efficiencies of 1.9%. Not only are better $V_{oc}$ (260 mV) and $J_{sc}$ (23 mAcm$^{-2}$) values achieved, but also the introduction of a dipole between the AZO electrode and the ZnO layer substantially increases the *FF* parameter, from 26% to 31%.

**Table 1**. Photovoltaic parameters of CZTSe solar cells with ZnO, PEI and ZnO/PEI as ETLs. Measurements were performed using standard test conditions (AM1.5G 1 kWm$^{-2}$, T = 25 ºC).

| DEVICE | ETL | $V_{oc}$ (mV) | $J_{sc}$ (mAcm$^{-2}$) | FF (%) | $\eta$ (%) |
|---|---|---|---|---|---|
| S01 | ZnO | 248 | 21 | 26 | 1.32 |
| S02 | PEI | 120 | 3.4 | 23 | 0.08 |
| S03 | ZnO/PEI | 260 | 23 | 31 | 1.9 |

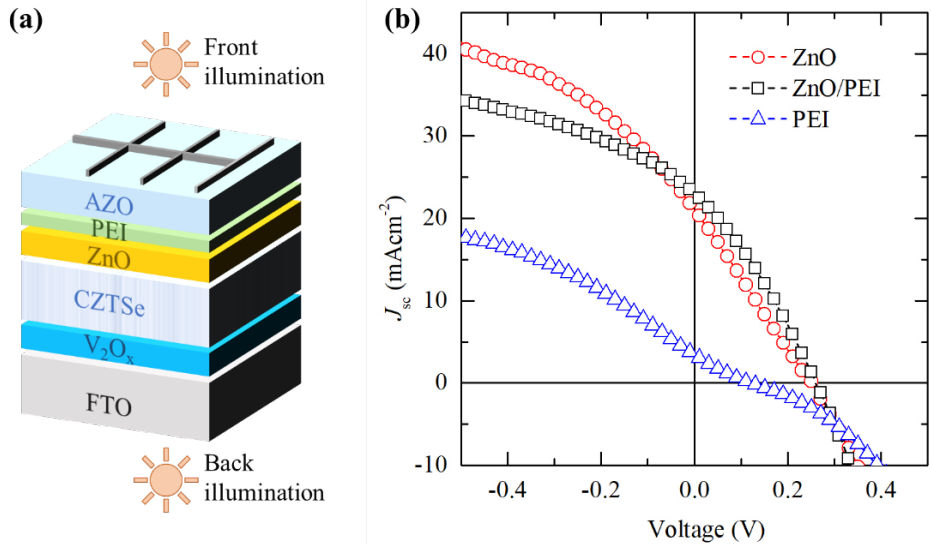

**Figure 1**. Sketch of the CZTSe solar cell (a) fabricated with ZnO/PEI as ETL (S03 structure). *J-V* characteristics (b) using front illumination for the three device structures using ZnO (S01), PEI (S02) and ZnO/PEI (S03) as ETL.

To elucidate this enhancement in the electron-selective behavior of the ZnO/PEI configuration, the plausible band diagram of complete device is discussed. Figure 2a shows the basic structure of a solar cell, which consist of an absorber layer with two selective contacts. The HTL blocks electrons (not allow them to recombine) and conduct holes to the external electrode (anode), while the ETL contact blocks holes (not allow them to recombine) and conduct electrons to the external electrode (cathode). Figure 2b shows the vacuum-aligned band diagram of CZTSe solar cells using the work function ($\phi$) and the bandgap reported values of $V_2O_x$, AZO and FTO from ref. [64], ZnO from ref. [65] and CZTSe from ref. [66], [67].

The inclusion of a dipole causes a displacement of the electrochemical potential of the materials connected with it, like applying a voltage [68], see Figure 2c. Therefore, using a PEI film between ZnO and AZO layers makes a work function reduction ($q\Delta$) of the AZO film, resulting in an apparent reduced work function of the TCO electrode, $\phi'_{AZO} = \phi_{AZO} - q\Delta$ [43].

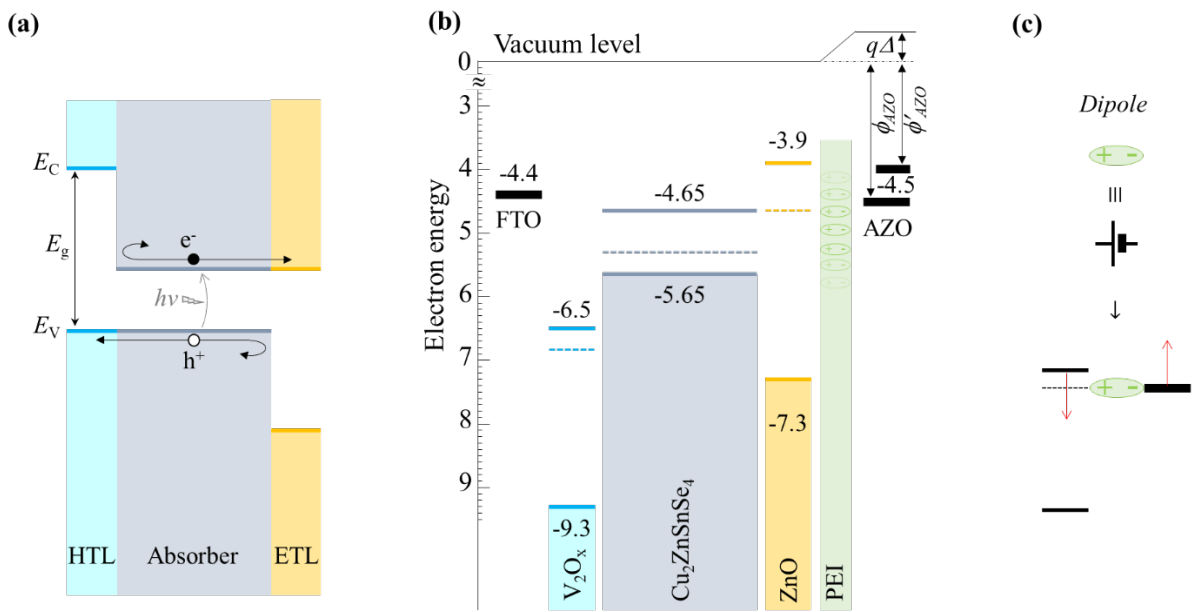

**Figure 2**. Simplified solar cell structure (a) in which photogenerated carriers are selectively collected. Vacuum-aligned band diagram of CZTSe solar cell based on $V_2O_x$ (HTL) and ZnO (ETL) as carrier-selective contacts (b). Illustration of the dipole effect using electrochemical potentials (c).

A schematic of the possible alignment of the energy bands for the ETL structure without and with the intercalation of the PEI dipole is shown in Figures 3a and 3b. When the electrochemical potentials of kesterite and ZnO balance, a downward curvature of the conduction and valence bands appears in the kesterite, causing an accumulation of electrons on the absorber surface, as well as an energy barrier for the hole transport. Hole blocking in the contact is notably increased due to the heterojunction between the kesterite absorber and the ZnO layer. In this way, electrons can flow to the AZO electrode, reaching the external circuit through the ZnO layer, whereas holes cannot. However, as it can be seen in Figure 3a and Figure 3b, there is still a certain energy barrier for the electrons located in the kesterite, i.e. energy spike, to access the outer electrode.

The inclusion of the PEI layer between the ZnO layer and the AZO layer improves the electrical performance of the solar cell. The most plausible explanation for this improvement is that the intercalation of the layer with a dipole moment causes a mismatch of the electrochemical potentials of the ZnO layer and that of AZO, causing a reduction in the work function of the AZO. This phenomenon results in an accumulation of electrons and in a reduction of the energy barrier seen by the electrons located in the kesterite, which would improve the thermionic field emission [69], as it can be seen in Figure 3c. Summarizing, our hypothesis is that the inclusion of the dipole reduces the energy barrier for electrons, facilitating the flow of electrons from the kesterite to the AZO electrode, through the ZnO layer (see Figure 3c).

Finally, a band alignment of the HTL structure is also proposed (see Figure S2 in the Supporting Information). The mechanism of operation of the HTL layer is the same as that which has been reported for other devices based on crystalline silicon [70] or organic semiconductors [71]. The high work function of $V_2O_x$ causes a curvature of the energy bands (upwards) in the kesterite semiconductor. This band curvature causes a buildup of holes in the kesterite ($V_2O_x$/kesterite interface) and a large potential barrier for electrons to flow into the $V_2O_x$. Therefore, the photogenerated holes in the kesterite are able to recombine with the electrons from the external circuit through the $V_2O_x$ film, which is of note that it is an n-type semiconductor with a Fermi level close to the conduction band [71].

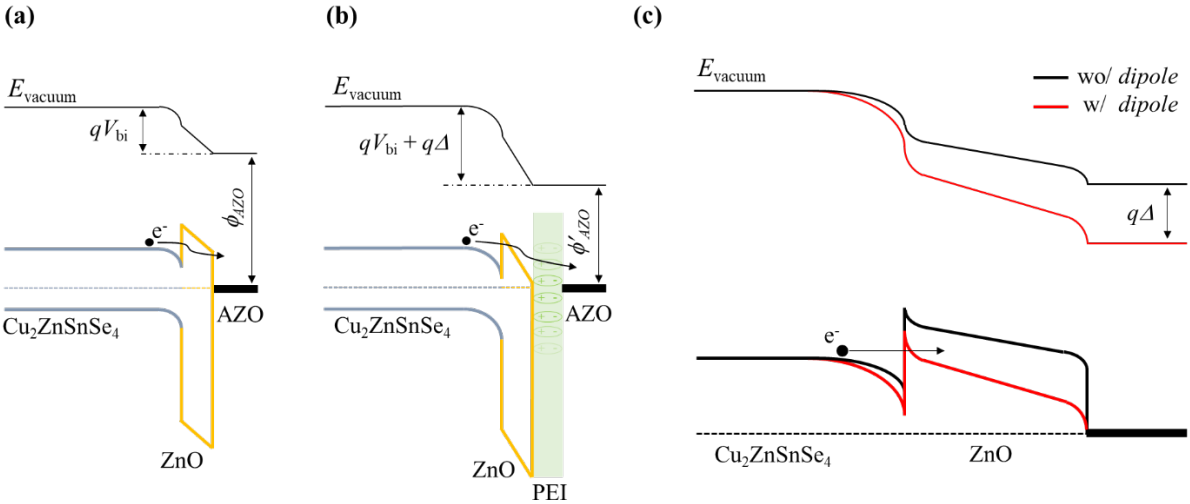

**Figure 3**. Proposed band alignment of ZnO (a) and ZnO/PEI (b) electron-selective contact. Zoom of the kesterite/ZnO interface with and without the use of dipole (c).

*B. Effect of ALD deposition temperature on solar cell performance:* In this section we analyzed the impact of the ALD deposition temperature on the ZnO layer the ETL electrical performance using the aforementioned S03 structure. Photovoltaic parameters considering different ALD deposition temperatures, i.e. 100, 110 and 125 ºC, are shown in Figure 4.

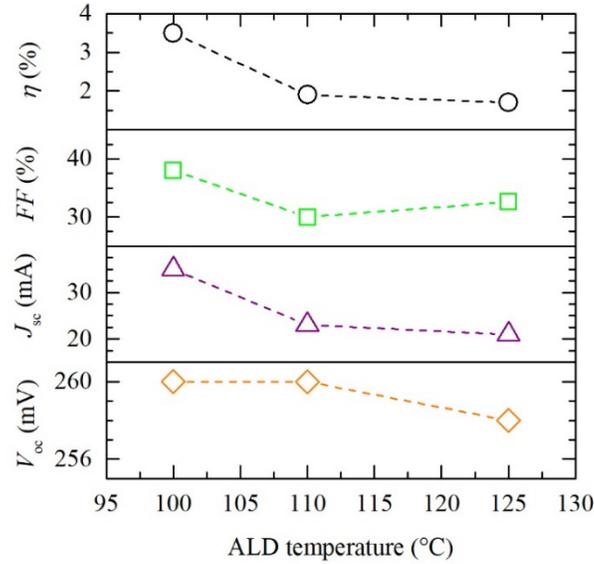

**Figure 4**. Parameters of solar cells for different ALD deposition temperatures.

It is clear the improvement of the solar cell efficiency when the deposition temperature decreases. The efficiency grows significantly from 1.7% to 3.5% when the deposition temperature decreases from 125 °C to 100 °C. This efficiency improvement is directly correlated with a better dark diode electrical behavior for lower deposition temperatures, as it can be seen in Figure 5. In fact, the sample processed at low temperature exhibits higher forward currents indicative of a lower series resistance. *FF* is in consequence the parameter which improves more at lower deposition temperatures. However, the $J_{sc}$ parameter follows the same trend of *FF*, as a consequence of a flattened *J-V* characteristic around 0 volts for the lower deposition temperature, i.e. short-circuit conditions. In this case, the $J_{sc}$ parameter is more like the expected device photocurrent $J_{ph}$. The origin of the improvement of the dark *J-V* characteristic when the ZnO deposition temperature decreases, i.e. lower series resistance, is not clear and a further deep study will be required in the future to explain this behavior.

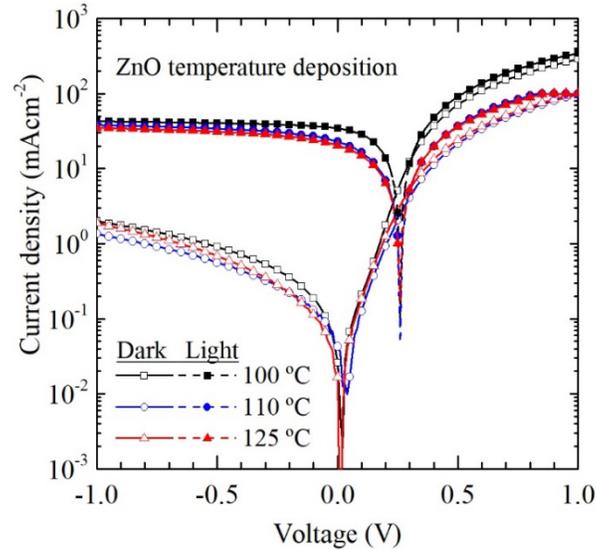

**Figure 5**. Dark and light semi-log *J-V* characteristics for the best solar cells with the ZnO layer deposited at different ALD temperatures, 100 ºC, 110 ºC and 125 ºC.

Regarding the best solar cell, i.e. ZnO deposited at 100 °C and S03 structure, the $J_{sc}$ value yields an outstanding value of ~35.0 mAcm$^{-2}$, even without using a light back reflector. The $V_{oc}$, *FF* and efficiency reach, 260 mV, 38% and 3.5%, respectively. The aforementioned parameters yield in average, 34 mAcm$^{-2}$, 250 mV, 35% and 3.1%, respectively, considering all devices fabricated in the same substrate (see Figure S3 in the Supporting Information).

Figure 6a shows the *EQE* response to the champion solar cell in two cases, front and back illumination. Results are directly compared with a conventional kesterite solar cell, i.e. opaque Mo-based HTL and CdS-based ETL. Regarding front illumination results, by replacing the standard ETL with the Cd-free approach, the parasitic absorption related to the CdS layer is effectively avoided at wavelengths below 600 nm (UV and a significant part of the visible spectrum) [29], [72]. The absorption peak at 460 nm reaches approximately 90% in the Cd-free device, while the response remains stable up to 1000 nm. Beyond 1000 nm, *EQE* curve starts to drop due to light absorption losses. It is worth mentioning that high *EQE* values about ~80% at 1000 nm wavelengths and non-zero back illumination *EQE* response can be explained by both the good passivation related to the $V_2O_x$-based HTL at the rear side and the relatively good behavior of CZTSe material as light absorber, i.e. a high effective diffusion lengths considering both bulk and rear surface recombination mechanisms [73]. Passivating the Kesterite surface has been reported as a crucial step to achieve a significant improvement in the efficiency of CZTSe solar cells and to resolve the absorber defects that cause a $V_{oc}$ deficit, as it is reported in previous studies [74]. Surprisingly, the conventional cell exhibits a poorer *EQE* performance than the Cd-free device in the 700 – 1300 nm wavelength range, which might be due to sputtering damage, indium contamination during ITO deposition at the last stage of its fabrication process [75], [76], [77] and/or due to a poor rear surface passivation provided by the MoSe/Mo HTL. Regardless of recombination mechanisms (bulk or surface), a lower effective diffusion length [73] in the conventional solar cell could explain the poorer *EQE* response in the IR part of the spectrum. Integrating the *EQE* curve with the AM1.5G solar spectrum yield generated photocurrents of ~35 mAcm$^{-2}$ and ~5 mAcm$^{-2}$ for the non-conventional cell under front and back illumination, respectively, which confirms its bifacial behavior.

Figure 6b shows the *J-V* characteristics, using both front and back illumination using standard conditions (AM1.5G 1 kWm$^{-2}$ solar spectrum, T = 25 ºC) for the best solar cell in comparison to a conventional device. Table 2, summarizes photovoltaic measurements for these cells. The *J-V* measurements confirm the remarkable $J_{sc}$ value achieved by the Cd-free structure, reaching 34.9 mAcm$^{-2}$ under front illumination compared to the 25.2 mAcm$^{-2}$ achieved by the CdS-based solar cell; both values are in agreement with the *EQE* measurements. It is important to mention that, although having a lower $V_{oc}$, the Cd-free structure achieves similar efficiency than the conventional counterpart, 3.5% and 3.4%, respectively, with the main additional advantage of not using Cd compounds in its fabrication process. These results demonstrate the high potential of kesterite solar cells based on Cd-free transparent TMO materials as carrier selective contacts.

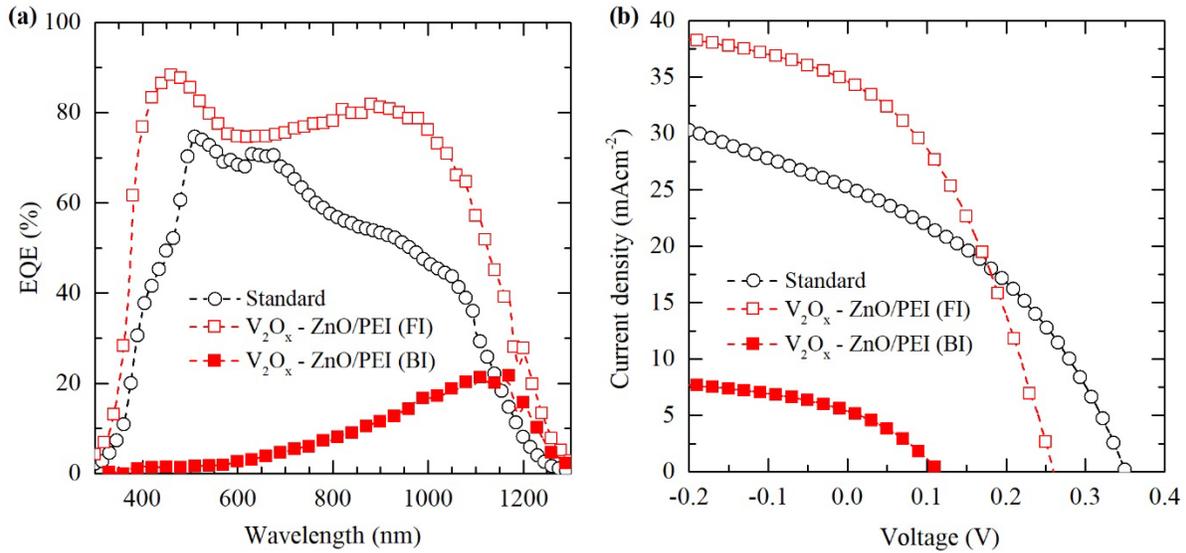

**Figure 6**. External quantum efficiency (EQE) curves (a) and *J-V* characteristics (b) for the champion Cd-free CZTSe solar cell measured under front and back illumination. Results are directly compared with a conventional kesterite solar cell under front illumination.

**Table 2**: Electrical characteristics of the best fabricated solar cell in comparison to a conventional structure using standard test conditions (AM1.5G 1 kWm$^{-2}$ solar spectrum and T = 25 ºC). Samples do not have any post-annealing process.

| Solar cell Structure | Light | $V_{oc}$ (mV) | $J_{sc}$ (mAcm$^{-2}$) | FF (%) | $\eta$ (%) |
|---|---|---|---|---|---|
| FTO/V$_2$O$_x$/CZTSe/ZnO/PEI/AZO | Front | 260 | 34.9 | 38.0 | 3.5 |
|  | Back | 160 | 5.3 | 35.8 | 0.3 |
| Mo/MoSe/CZTSe/CdS/ZnO/ITO | Front | 349 | 25.2 | 38.1 | 3.4 |

*C. Optical Analysis of ETL and HTL stacks*: In order to study optical properties of Glass/ZnO/PEI/AZO based ETL and Glass/FTO/$V_2O_x$ based HTL stacks UV-Visible-NIR spectroscopy measurements in the 300 – 2000 nm wavelength range were made.

To evaluate the impact of the PEI dipole in the solar cell absorption, transmittance and reflectance measurements of ZnO/AZO stacks, with and without dipole interlayer, were performed, and results are presented through the Figure 7a. The inclusion of PEI dipole interlayer produces an imperceptible change in the transmission and reflectance curve. Figure 7b shows the absorption characteristic for the ZnO/PEI/AZO-based ETL as compared to ZnO/AZO. The absorption of the glass is also shown in Figure 7b, which is practically insignificant and has no influence on the measurements. In conclusion, both structures, with and without PEI, allow the passage of light with minimal absorption. From an optical point of view, the replacement of the absorbent CdS layer with a transparent TMO-based ETL contact is feasible for improving the kesterite solar cell performance [78].

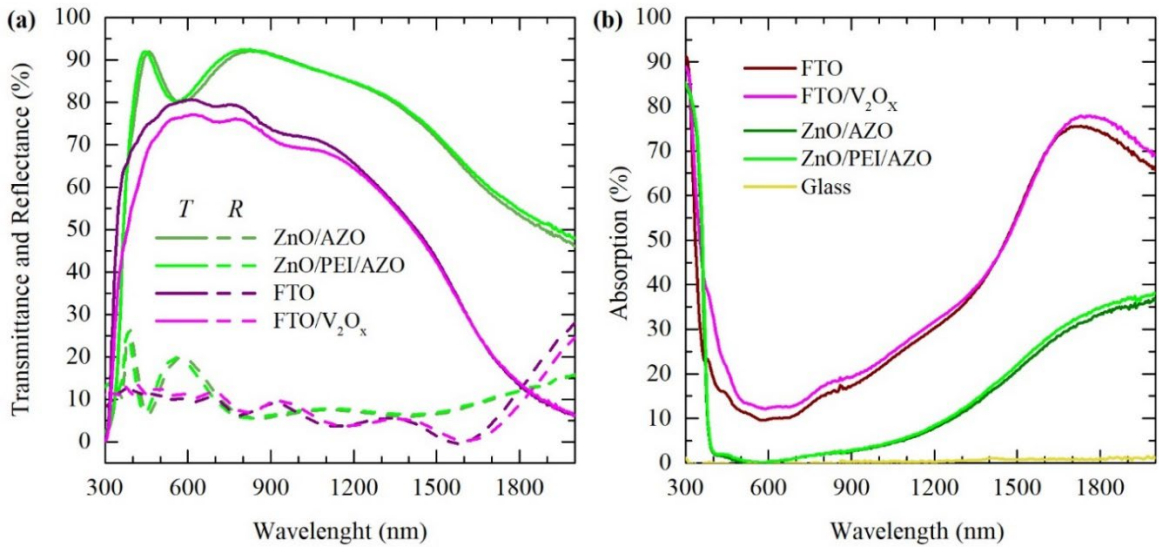

**Figure 7**. Transmittance and reflectance (a) and absorption (b) *vs.* wavelength curves considering Glass/ZnO/PEI/AZO, Glass/ZnO/AZO, Glass/FTO and Glass/FTO/$V_2O_x$ stacks. The absorption of a bare glass substrate is also depicted in (b).

Figure 7a also shows the transmittance and reflectance curve of the back contact (Glass/FTO/$V_2O_x$) in comparison to a sample without $V_2O_x$ (Glass/FTO). The FTO layer unfortunately contributes to the significant optical losses. As it can be seen in Figure 7b, the presence of the FTO layer accounts for around 20% of light absorption in the whole spectrum under study, whereas it is almost negligible in the $V_2O_x$ film. Therefore, the FTO electrode impacts the cell performance in different ways depending on which surface is illuminated, front or back illumination. On the one hand, using front illumination, only the FTO absorption at long wavelengths penalizes scarcely the *EQE* response in the NIR part of the spectrum, i.e. photons not absorbed in the kesterite material which arrive to the back electrode. It is worth mentioning that this phenomenon will be more prominent for the thinner absorber layers. On the other hand, by using rear illumination the FTO absorption plays, a crucial role in this case, by substantially decreasing the *EQE* response in the whole wavelength range.

*D. Compositional and structural analysis*: Structural and compositional analysis were performed by means of STEM images and EDS analysis respectively using a S03 CZTSe solar cell. Figure 8 shows STEM images which correspond to cross-section views of the whole device. The total device thickness is about 2.2 µm considering the FTO layer. As it is observed in Figure 8, the growth of the kesterite absorber onto the $V_2O_x$ layer is enough conformal and compact, although dark areas are observed which represent void zones. These structural defects are present in other CZTSe structures based on Mo as back contact [79], and they could be the result of the grains growth with different orientation in the absorber. Except for these located defects, the kesterite absorber has good morphology, the crystal grains are medium/large size and the CZTSe thickness is about ~1.4 µm. The EDS mapping images give evidence of the presence of the compositional elements of the absorber: Cu, Zn, Sn and Se. The oxygen compound is the whole structure, although it has low presence in the kesterite, it is present especially in the FTO and $V_2O_x$ layers in the back electrode, and in the ZnO and AZO layers in the front contact.

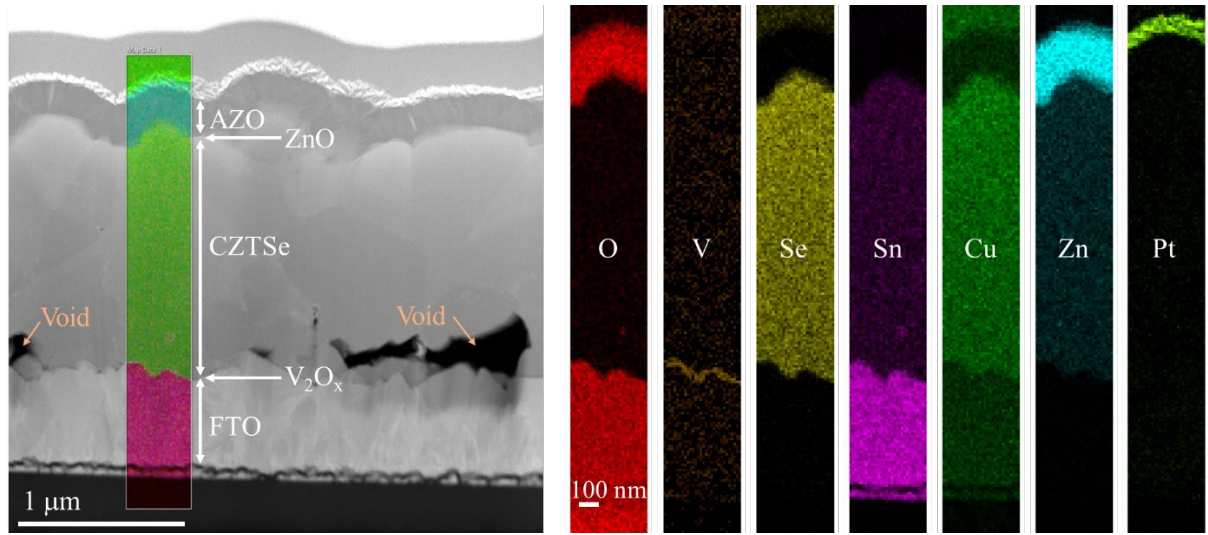

**Figure 8**. STEM Image of Cross-sectional view of CZTSe solar cell and EDS mapping images.

Figure 9 shows a cross-section close view STEM image of the ETL contact formed by ZnO/PEI/AZO stack. The EDS mapping show a clear view of the elemental composition of these layers, confirming the presence of O, Zn and Al elements in the stack. However, the Al element exists apparently in the whole structure, including the kesterite substrate. The ZnO layer can be differentiated due to the low density of Al in the edge near the kesterite material. According to the STEM image, the AZO and ZnO layers have a thickness of about 200 nm and 20 nm, respectively. Unfortunately, the ultra-thin PEI film has not been recognized either under STEM images or EDS analysis due to the resolution of the images.

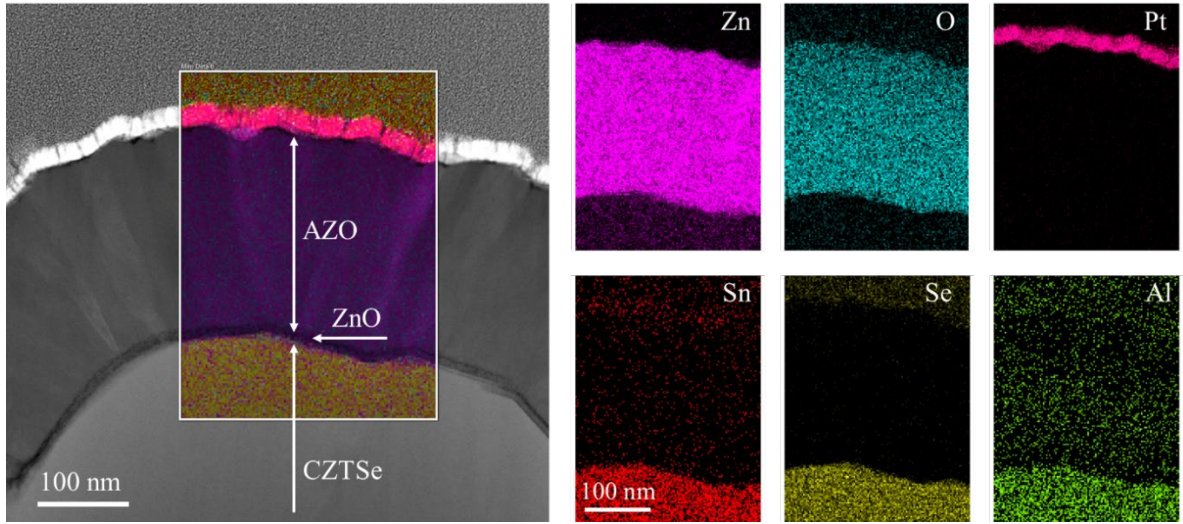

**Figure 9**. STEM Image of a cross-section view of the ETL stack and EDS mapping images.

Figure 10 shows the cross-section STEM image and EDS compositional mappings of the FTO/$V_2O_x$ based back electrode. As it can be deduced in Figure 10, the $V_2O_x$ layer has a thickness of about 10 nm. Oxygen is present in whole FTO/$V_2O_x$ stack as expected. The vanadium element is located on the boundary with the kesterite, showing a conformal and uniform growth on the FTO. The presence of Cu in the stack is due to the diffusion of this element from the absorber towards the whole structure. This phenomenon is already reported in other studies [80]. Notice that Se and Zn remains in the absorber showing a low diffusion of these elements to the rest of the device layers. The FTO film has a high presence of Sn and O elements as expected.

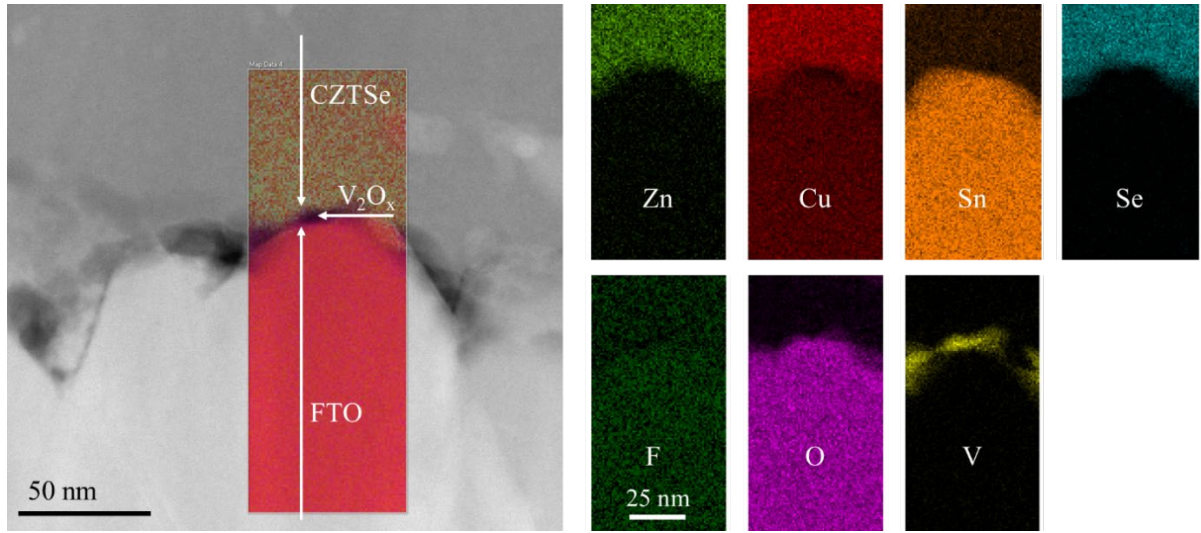

**Figure 10**. STEM Cross-section view and EDS mapping images of the CZTSe solar cell back contact.

CONCLUSIONS

In this work, TMOs were studied as carrier selective contacts for the CZTSe kesterite solar cells. The ETL and HTL contacts were based on transparent ZnO/PEI/AZO and $V_2O_x$/FTO stacks, respectively. In our approach the ZnO, AZO and $V_2O_x$ layers were deposited by atomic layer deposition technique. Using these contact schemes, CdS-free bifacial solar cells were fabricated demonstrating outstanding short-circuit current density of 35 mAcm$^{-2}$ with photovoltaic efficiencies of about 3.5% (under front illumination). Using back illumination, these parameters value were 5.3 mAcm$^{-2}$ and 0.3%, respectively. These values clearly indicate bifacial nature of the fabricated solar cell devices though the performance is mainly limited due to get parasitic light absorption in the FTO. The use of a PEI dipole interlayer boosts the fill factor of the solar cell improving the contact resistivity of the electron-selective contact by improving the band alignment between the absorber and the TCO. The hole-selective contact based on $V_2O_x$ improves the rear passivation of the device and allows a conformal and uniform growth of the absorber onto glass/FTO substrates as a replacement of glass/Mo-based substrates. From optical analysis of the contact layers, peak transmittances of ~90% and ~80% are achieved for the front and back contact schemes, respectively. The proposed ETL contact avoids completely the parasitic light absorption introduced by the CdS layer used in conventional CZTSe solar cells. This novel approach opens the door to the fabrication of Cd-free bifacial thin-film solar cells for BIPV applications, such as agrivoltaics where some transparency is required. Moreover, the ability to collect the light from both device sides is a key factor to boost photovoltaic efficiency in single and tandem PV technologies.

## ASSOCIATED CONTENT

Supporting Information is available from the author in another attached file.

## DECLARATION OF COMPETING INTEREST

The authors declare that they have no known competing financial interests or personal relationships that could have appeared to influence the work reported in this paper.

## AUTHOR INFORMATION

**Corresponding Author**

R. Almache-Hernández*: ralmache@yachaytech.edu.ec

G. Masmitjà*: gerard.masmitja@upc.edu

**Author Contributions**

The manuscript was written through contributions of all authors. All authors have given approval to the final version of the manuscript. ‡ These authors contributed equally.

ACKNOWLEDGMENT

Authors acknowledge the use of instrumentation as well as the technical advice provided by the National Facility ELECMI ICTS: Laboratorio de Microscopias Avanzadas at Universidad de Zaragoza. This work has been supported by the Ecuadorian government grants SENESCYT (BECAS DE POSGRADOS INTERNACIONALES 2018) and the Spanish government under projects PID2022-138434OB-C51 (SCALING), PID2020-115719RB-C21 (GETPV), TED2021-131778B-C22 (TROPIC) and TED2021-129758B-C32 (TransEl) funded by MCIN/AEI/10.13039/501100011033. TROPIC and TransEl projects have also support from the European Union "NextGenerationEU"/PRTR program.

REFERENCES

[1] M. V. Dambhare, B. Butey, and S. V. Moharil, "Solar photovoltaic technology: A review of different types of solar cells and its future trends," *J Phys Conf Ser*, vol. 1913, no. 1, p. 012053, May 2021, doi: 10.1088/1742-6596/1913/1/012053.

[2] V. A. Milichko *et al.*, "Solar photovoltaics: current state and trends," *Physics-Uspekhi*, vol. 59, no. 8, p. 727, Aug. 2016, doi: 10.3367/UFNE.2016.02.037703.

[3] R. W. Miles, "Photovoltaic solar cells: Choice of materials and production methods," *Vacuum*, vol. 80, no. 10, pp. 1090–1097, Aug. 2006, doi: 10.1016/J.VACUUM.2006.01.006.

[4] A. G. Aberle, "Thin-film solar cells," *Thin Solid Films*, vol. 517, no. 17, pp. 4706–4710, Jul. 2009, doi: 10.1016/J.TSF.2009.03.056.

[5] M. Yamaguchi, K. H. Lee, K. Araki, and N. Kojima, "A review of recent progress in heterogeneous silicon tandem solar cells," *J Phys D Appl Phys*, vol. 51, no. 13, p. 133002, Mar. 2018, doi: 10.1088/1361-6463/AAAF08.

[6] M. A. Green, A. W. Blakers, S. Narayanan, and M. Taouk, "Improvements in silicon solar cell efficiency," *Solar Cells*, vol. 17, no. 1, pp. 75–83, Mar. 1986, doi: 10.1016/0379-6787(86)90060-8.

[7] A. Shah, P. Torres, R. Tscharner, N. Wyrsch, and H. Keppner, "Photovoltaic Technology: The Case for Thin-Film Solar Cells," *Science (1979)*, vol. 285, no. 5428, pp. 692–698, Jul. 1999, doi: 10.1126/SCIENCE.285.5428.692.

[8]     V. Fthenakis, "Sustainability of photovoltaics: The case for thin-film solar cells," *Renewable and Sustainable Energy Reviews*, vol. 13, no. 9, pp. 2746–2750, Dec. 2009, doi: 10.1016/J.RSER.2009.05.001.

[9]     N. L. Muttumthala and A. Yadav, "A concise overview of thin film photovoltaics," *Mater Today Proc*, vol. 64, pp. 1475–1478, Jan. 2022, doi: 10.1016/J.MATPR.2022.04.862.

[10]    S. Amiri and S. Dehghani, "Design of Highly Efficient CZTS/CZTSe Tandem Solar Cells," *J Electron Mater*, vol. 49, no. 3, pp. 2164–2172, Mar. 2020, doi: 10.1007/s11664-019-07898-w.

[11]    K. J. Tiwari, S. Giraldo, M. Placidi, Z. Jehl Li-Kao, and E. Saucedo, "Recent Advances in the Kesterite-Based Thin Film Solar Cell Technology: Role of Ge," pp. 41–66, 2022, doi: 10.1007/978-981-19-3724-8_3.

[12]    A. Wang, M. He, M. A. Green, K. Sun, and X. Hao, "A Critical Review on the Progress of Kesterite Solar Cells: Current Strategies and Insights," *Adv Energy Mater*, vol. 13, no. 2, p. 2203046, Jan. 2023, doi: 10.1002/AENM.202203046.

[13]    M. Espindola-Rodriguez *et al.*, "Efficient bifacial Cu2ZnSnSe4 solar cells," *2015 IEEE 42nd Photovoltaic Specialist Conference, PVSC 2015*, Dec. 2015, doi: 10.1109/PVSC.2015.7355903.

[14]    H. Katagiri, "Cu2ZnSnS4 thin film solar cells," *Thin Solid Films*, vol. 480–481, pp. 426–432, Jun. 2005, doi: 10.1016/J.TSF.2004.11.024.

[15]    K. L. Chopra, P. D. Paulson, and V. Dutta, "Thin-film solar cells: an overview," *Progress in Photovoltaics: Research and Applications*, vol. 12, no. 2–3, pp. 69–92, Mar. 2004, doi: 10.1002/PIP.541.

[16]    M. Edoff, "Thin film solar cells: Research in an industrial perspective," *Ambio*, vol. 41, no. SUPPL.2, pp. 112–118, Mar. 2012, doi: 10.1007/s13280-012-0265-6.

[17]    F. I. Lai, J. F. Yang, Y. L. Wei, and S. Y. Kuo, "High quality sustainable Cu2ZnSnSe4 (CZTSe) absorber layers in highly efficient CZTSe solar cells," *Green Chemistry*, vol. 19, no. 3, pp. 795–802, Feb. 2017, doi: 10.1039/C6GC02300B.

[18]    G. Zoppi, I. Forbes, R. W. Miles, P. J. Dale, J. J. Scragg, and L. M. Peter, "Cu2ZnSnSe4 thin film solar cells produced by selenisation of magnetron sputtered precursors," *Progress in Photovoltaics: Research and Applications*, vol. 17, no. 5, pp. 315–319, Aug. 2009, doi: 10.1002/PIP.886.

[19]    S. Delbos, "Kësterite thin films for photovoltaics : a review," *EPJ Photovoltaics*, vol. 3, p. 35004, 2012, doi: 10.1051/EPJPV/2012008.

[20]    M. Rehan *et al.*, "Fabrication and Characterization of Cu2ZnSnSe4 Thin-Film Solar Cells using a Single-Stage Co-Evaporation Method: Effects of Film Growth Temperatures on Device Performances," *Energies 2020, Vol. 13, Page 1316*, vol. 13, no. 6, p. 1316, Mar. 2020, doi: 10.3390/EN13061316.

[21]    Z. Zhang *et al.*, "Modified Back Contact Interface of CZTSe Thin Film Solar Cells: Elimination of Double Layer Distribution in Absorber Layer," *Advanced Science*, vol. 5, no. 2, p. 1700645, Feb. 2018, doi: 10.1002/ADVS.201700645.

[22] Y. S. Lee *et al.*, "Cu2ZnSnSe4 Thin-Film Solar Cells by Thermal Co-evaporation with 11.6% Efficiency and Improved Minority Carrier Diffusion Length," *Adv Energy Mater*, vol. 5, no. 7, p. 1401372, Apr. 2015, doi: 10.1002/AENM.201401372.

[23] G. M. Ilari, C. M. Fella, C. Ziegler, A. R. Uhl, Y. E. Romanyuk, and A. N. Tiwari, "Cu2ZnSnSe4 solar cell absorbers spin-coated from amine-containing ether solutions," *Solar Energy Materials and Solar Cells*, vol. 104, pp. 125–130, Sep. 2012, doi: 10.1016/J.SOLMAT.2012.05.004.

[24] Y. Li *et al.*, "Suppressing Element Inhomogeneity Enables 14.9% Efficiency CZTSSe Solar Cells," *Advanced Materials*, p. 2400138, 2024, doi: 10.1002/ADMA.202400138.

[25] S. Giraldo, Z. Jehl, M. Placidi, V. Izquierdo-Roca, A. Pérez-Rodríguez, and E. Saucedo, "Progress and Perspectives of Thin Film Kesterite Photovoltaic Technology: A Critical Review," *Advanced Materials*, vol. 31, no. 16, p. 1806692, Apr. 2019, doi: 10.1002/ADMA.201806692.

[26] O. Almora, L. G. Gerling, C. Voz, R. Alcubilla, J. Puigdollers, and G. Garcia-Belmonte, "Superior performance of V2O5 as hole selective contact over other transition metal oxides in silicon heterojunction solar cells," *Solar Energy Materials and Solar Cells*, vol. 168, pp. 221–226, Aug. 2017, doi: 10.1016/J.SOLMAT.2017.04.042.

[27] G. Masmitjà *et al.*, "Interdigitated back-contacted crystalline silicon solar cells fully manufactured with atomic layer deposited selective contacts," *Solar Energy Materials and Solar Cells*, vol. 240, p. 111731, Jun. 2022, doi: 10.1016/J.SOLMAT.2022.111731.

[28] I. Becerril-Romero *et al.*, "Transition-Metal Oxides for Kesterite Solar Cells Developed on Transparent Substrates," *ACS Appl Mater Interfaces*, vol. 12, no. 30, pp. 33656–33669, Jul. 2020, doi: 10.1021/acsami.0c06992.

[29] G. Tseberlidis *et al.*, "Titania as Buffer Layer for Cd-Free Kesterite Solar Cells," *ACS Mater Lett*, vol. 5, no. 1, pp. 219–224, Jan. 2023, doi: 10.1021/ACSMATERIALSLETT.2C00933.

[30] C. Gobbo *et al.*, "Effect of the ZnSnO/AZO Interface on the Charge Extraction in Cd-Free Kesterite Solar Cells," *Energies 2023, Vol. 16, Page 4137*, vol. 16, no. 10, p. 4137, May 2023, doi: 10.3390/EN16104137.

[31] Z. Zhang *et al.*, "Over 10% Efficient Pure CZTSe Solar Cell Fabricated by Electrodeposition with Ge Doping," *Solar RRL*, vol. 4, no. 5, p. 2000059, May 2020, doi: 10.1002/SOLR.202000059.

[32] H. S. Nugroho *et al.*, "A progress review on the modification of CZTS(e)-based thin-film solar cells," *Journal of Industrial and Engineering Chemistry*, vol. 105, pp. 83–110, Jan. 2022, doi: 10.1016/J.JIEC.2021.09.010.

[33] M. Nadarajah, K. S. Gour, and V. N. Singh, "Sputtered Cadmium Sulfide (CdS) Buffer Layer for Kesterite and Chalcogenide Thin Film Solar Cell (TFSC) Applications," *J Nanosci Nanotechnol*, vol. 20, no. 6, pp. 3909–3912, Nov. 2019, doi: 10.1166/JNN.2020.17528.

[34] T. Nagai *et al.*, "Electronic structures of Cu2ZnSnSe4 surface and CdS/Cu2ZnSnSe4 heterointerface," *Jpn J Appl Phys*, vol. 56, no. 6, p. 065701, Jun. 2017, doi: 10.7567/JJAP.56.065701.

[35]    M. He, K. Sun, M. P. Suryawanshi, J. Li, and X. Hao, "Interface engineering of p-n heterojunction for kesterite photovoltaics: A progress review," *Journal of Energy Chemistry*, vol. 60, pp. 1–8, Sep. 2021, doi: 10.1016/J.JECHEM.2020.12.019.

[36]    S. Tripathi, Sadanand, P. Lohia, and D. K. Dwivedi, "Contribution to sustainable and environmental friendly non-toxic CZTS solar cell with an innovative hybrid buffer layer," *Solar Energy*, vol. 204, pp. 748–760, Jul. 2020, doi: 10.1016/J.SOLENER.2020.05.033.

[37]    X. Liu *et al.*, "The current status and future prospects of kesterite solar cells: a brief review," *Progress in Photovoltaics: Research and Applications*, vol. 24, no. 6, pp. 879–898, Jun. 2016, doi: 10.1002/PIP.2741.

[38]    Q. An *et al.*, "High performance planar perovskite solar cells by ZnO electron transport layer engineering," *Nano Energy*, vol. 39, pp. 400–408, Sep. 2017, doi: 10.1016/J.NANOEN.2017.07.013.

[39]    Z. Ma, Z. Tang, E. Wang, M. R. Andersson, O. Inganäs, and F. Zhang, "Influences of surface roughness of ZnO electron transport layer on the photovoltaic performance of organic inverted solar cells," *Journal of Physical Chemistry C*, vol. 116, no. 46, pp. 24462–24468, Nov. 2012, doi: 10.1021/jp308480u.

[40]    S. Yang *et al.*, "Sulfurizing Sputtered-ZnO as buffer layer for cadmium-free Cu2ZnSnS4 solar cells," *Mater Sci Semicond Process*, vol. 101, pp. 87–94, Oct. 2019, doi: 10.1016/J.MSSP.2019.05.021.

[41]    H. K. Hong, G. Y. Song, H. J. Shim, J. H. Kim, and J. Heo, "Recovery of rectifying behavior in Cu2ZnSn(S,Se)4/Zn(O,S) thin-film solar cells by in-situ nitrogen doping of buffer layers," *Solar Energy*, vol. 145, pp. 20–26, Mar. 2017, doi: 10.1016/J.SOLENER.2016.09.042.

[42]    L. Yan, Y. Song, Y. Zhou, B. Song, and Y. Li, "Effect of PEI cathode interlayer on work function and interface resistance of ITO electrode in the inverted polymer solar cells," *Org Electron*, vol. 17, pp. 94–101, Feb. 2015, doi: 10.1016/J.ORGEL.2014.11.023.

[43]    D. Rovira *et al.*, "Polymeric Interlayer in CdS-Free Electron-Selective Contact for Sb2Se3 Thin-Film Solar Cells," *Int J Mol Sci*, vol. 24, no. 4, p. 3088, Feb. 2023, doi: 10.3390/IJMS24043088/S1.

[44]    E. Ros *et al.*, "Expanding the Perspective of Polymeric Selective Contacts in Photovoltaic Devices Using Branched Polyethylenimine," *ACS Appl Energy Mater*, vol. 5, no. 9, pp. 10702–10709, Sep. 2022, doi: 10.1021/acsaem.2c01422.

[45]    F. C. Wu, K. C. Tung, W. Y. Chou, F. C. Tang, and H. L. Cheng, "Charge selectivity in polymer:Fullerene-based organic solar cells with a chemically linked polyethylenimine interlayer," *Org Electron*, vol. 29, pp. 120–126, Feb. 2016, doi: 10.1016/J.ORGEL.2015.11.037.

[46]    D. Yang, P. Fu, F. Zhang, N. Wang, J. Zhang, and C. Li, "High efficiency inverted polymer solar cells with room-temperature titanium oxide/polyethylenimine films as electron transport layers," *J Mater Chem A Mater*, vol. 2, no. 41, pp. 17281–17285, Sep. 2014, doi: 10.1039/C4TA03838J.

[47]    M. Buffière *et al.*, "Microstructural analysis of 9.7% efficient Cu2ZnSnSe4 thin film solar cells," *Appl Phys Lett*, vol. 105, no. 18, Nov. 2014, doi: 10.1063/1.4901401/132927.

[48]  L. Yao *et al.*, "The formation of MoSe2 films during selenization process in CZTSe solar cells," pp. 1–2, Oct. 2016, doi: 10.1109/INEC.2016.7589329.

[49]  R. Almache-Hernández *et al.*, "Hole Transport Layer based on atomic layer deposited V2Ox films: Paving the road to semi-transparent CZTSe solar cells," *Solar Energy*, vol. 226, pp. 64–71, Sep. 2021, doi: 10.1016/J.SOLENER.2021.08.007.

[50]  T. Nakada, Y. Hirabayashi, T. Tokado, D. Ohmori, and T. Mise, "Novel device structure for Cu(In,Ga)Se2 thin film solar cells using transparent conducting oxide back and front contacts," *Solar Energy*, vol. 77, no. 6, pp. 739–747, Dec. 2004, doi: 10.1016/J.SOLENER.2004.08.010.

[51]  A. Ruiz-Perona *et al.*, "Effect of Na and the back contact on Cu2Zn(Sn,Ge)Se4 thin-film solar cells: Towards semi-transparent solar cells," *Solar Energy*, vol. 206, pp. 555–563, Aug. 2020, doi: 10.1016/J.SOLENER.2020.06.044.

[52]  J. S. Kim, J. K. Kang, and D. K. Hwang, "High efficiency bifacial Cu2ZnSnSe4 thin-film solar cells on transparent conducting oxide glass substrates," *APL Mater*, vol. 4, no. 9, Sep. 2016, doi: 10.1063/1.4962145/121515.

[53]  M. Espindola-Rodriguez *et al.*, "Bifacial Kesterite Solar Cells on FTO Substrates," *ACS Sustain Chem Eng*, vol. 5, no. 12, pp. 11516–11524, Dec. 2017, doi: 10.1021/acssuschemeng.7b02797.

[54]  S. Temgoua, R. Bodeux, F. Mollica, and N. Naghavi, "Comparative study of Cu2ZnSnSe4 solar cells growth on transparent conductive oxides and molybdenum substrates," *Solar Energy*, vol. 194, pp. 121–127, Dec. 2019, doi: 10.1016/J.SOLENER.2019.10.050.

[55]  F. I. Lai, J. F. Yang, W. C. Chen, Y. C. Hsu, and S. Y. Kuo, "Weatherability of Cu2ZnSnSe4 thin film solar cells on diverse substrates," *Solar Energy*, vol. 195, pp. 626–635, Jan. 2020, doi: 10.1016/J.SOLENER.2019.11.089.

[56]  E. R. Costals *et al.*, "Atomic layer deposition of vanadium oxide films for crystalline silicon solar cells," *Mater Adv*, vol. 3, no. 1, pp. 337–345, Jan. 2022, doi: 10.1039/D1MA00812A.

[57]  R. Raj, H. Gupta, and L. P. Purohit, "Performance of V2O5 hole selective layer in CdS/CdTe heterostructure solar cell," *J Alloys Compd*, vol. 907, p. 164408, Jun. 2022, doi: 10.1016/J.JALLCOM.2022.164408.

[58]  S. A. Awuku, F. Muhammad-Sukki, and N. Sellami, "Building Integrated Photovoltaics—The Journey So Far and Future," *Energies 2022, Vol. 15, Page 1802*, vol. 15, no. 5, p. 1802, Feb. 2022, doi: 10.3390/EN15051802.

[59]  S. Gorjian *et al.*, "Progress and challenges of crop production and electricity generation in agrivoltaic systems using semi-transparent photovoltaic technology," *Renewable and Sustainable Energy Reviews*, vol. 158, p. 112126, Apr. 2022, doi: 10.1016/J.RSER.2022.112126.

[60]  H. Deng *et al.*, "Novel symmetrical bifacial flexible CZTSSe thin film solar cells for indoor photovoltaic applications," *Nature Communications 2021 12:1*, vol. 12, no. 1, pp. 1–8, May 2021, doi: 10.1038/s41467-021-23343-1.

[61] E. Aydin *et al.*, "Sputtered transparent electrodes for optoelectronic devices: Induced damage and mitigation strategies," *Matter*, vol. 4, no. 11, pp. 3549–3584, Nov. 2021, doi: 10.1016/J.MATT.2021.09.021.

[62] T. Tynell and M. Karppinen, "Atomic layer deposition of ZnO: a review," *Semicond Sci Technol*, vol. 29, no. 4, p. 043001, Feb. 2014, doi: 10.1088/0268-1242/29/4/043001.

[63] V. Quemener *et al.*, "The work function of n-ZnO deduced from heterojunctions with Si prepared by ALD," *J Phys D Appl Phys*, vol. 45, no. 31, p. 315101, Jul. 2012, doi: 10.1088/0022-3727/45/31/315101.

[64] Y. Ko, H. J. Park, C. Lee, Y. Kang, and Y. Jun, "Recent Progress in Interconnection Layer for Hybrid Photovoltaic Tandems," *Advanced Materials*, vol. 32, no. 51, p. 2002196, Dec. 2020, doi: 10.1002/ADMA.202002196.

[65] V. Quemener *et al.*, "The work function of n-ZnO deduced from heterojunctions with Si prepared by ALD," *J Phys D Appl Phys*, vol. 45, no. 31, p. 315101, Jul. 2012, doi: 10.1088/0022-3727/45/31/315101.

[66] L. A. Burton and A. Walsh, "Band alignment in SnS thin-film solar cells: Possible origin of the low conversion efficiency," *Appl Phys Lett*, vol. 102, no. 13, Apr. 2013, doi: 10.1063/1.4801313/125079.

[67] A. Walsh, S. Chen, S. H. Wei, and X. G. Gong, "Kesterite Thin-Film Solar Cells: Advances in Materials Modelling of Cu2ZnSnS4," *Adv Energy Mater*, vol. 2, no. 4, pp. 400–409, Apr. 2012, doi: 10.1002/AENM.201100630.

[68] J. Puigdollers, C. Voz, and E. Ros, "Physics and Technology of Carrier Selective Contact Based Heterojunction Silicon Solar Cells," *Energy Systems in Electrical Engineering*, vol. Part F2140, pp. 61–95, 2022, doi: 10.1007/978-981-19-4526-7_2.

[69] M. Garbrecht *et al.*, "Thermally stable epitaxial ZrN/carrier-compensated Sc0.99Mg0.01N metal/semiconductor multilayers for thermionic energy conversion," *J Mater Sci*, vol. 55, no. 4, pp. 1592–1602, Feb. 2020, doi: 10.1007/S10853-019-04127-X/METRICS.

[70] L. G. Gerling *et al.*, "Transition metal oxides as hole-selective contacts in silicon heterojunctions solar cells," *Solar Energy Materials and Solar Cells*, vol. 145, pp. 109–115, Feb. 2016, doi: 10.1016/J.SOLMAT.2015.08.028.

[71] M. T. Greiner, M. G. Helander, W. M. Tang, Z. Bin Wang, J. Qiu, and Z. H. Lu, "Universal energy-level alignment of molecules on metal oxides," *Nature Materials 2011 11:1*, vol. 11, no. 1, pp. 76–81, Nov. 2011, doi: 10.1038/nmat3159.

[72] J. Li *et al.*, "Tailoring the defects and carrier density for beyond 10% efficient CZTSe thin film solar cells," *Solar Energy Materials and Solar Cells*, vol. 159, pp. 447–455, Jan. 2017, doi: 10.1016/J.SOLMAT.2016.09.034.

[73] H. Fujiwara, A. Nakane, D. Murata, H. Tampo, T. Matsui, and H. Shibata, "Analysis of Optical and Recombination Losses in Solar Cells," *Springer Series in Optical Sciences*, vol. 214, pp. 29–82, 2018, doi: 10.1007/978-3-319-95138-6_2.

[74] J. Kim, S. Park, S. Ryu, J. Oh, and B. Shin, "Improving the open-circuit voltage of Cu2ZnSnSe4 thin film solar cells via interface passivation," *Progress in Photovoltaics: Research and Applications*, vol. 25, no. 4, pp. 308–317, Apr. 2017, doi: 10.1002/PIP.2864.

[75] Z. Su *et al.*, "Device Postannealing Enabling over 12% Efficient Solution-Processed Cu2ZnSnS4 Solar Cells with Cd2+ Substitution," *Advanced Materials*, vol. 32, no. 32, p. 2000121, Aug. 2020, doi: 10.1002/ADMA.202000121.

[76] M. Rehan *et al.*, "Defect Engineering in Earth-Abundant Cu2ZnSnSe4 Absorber Using Efficient Alkali Doping for Flexible and Tandem Solar Cell Applications," *Energy & Environmental Materials*, vol. 7, no. 2, p. e12604, Mar. 2024, doi: 10.1002/EEM2.12604.

[77] Y. Ji *et al.*, "An ITO-Free Kesterite Solar Cell," *Small*, vol. 20, no. 6, p. 2307242, Feb. 2024, doi: 10.1002/SMLL.202307242.

[78] N. E. Gorji, "Quantitative analysis of the optical losses in CZTS thin-film semiconductors," *IEEE Trans Nanotechnol*, vol. 13, no. 4, pp. 743–748, Jul. 2014, doi: 10.1109/TNANO.2014.2318057.

[79] R. Fonoll-Rubio *et al.*, "Insights into interface and bulk defects in a high efficiency kesterite-based device," *Energy Environ Sci*, vol. 14, no. 1, pp. 507–523, Jan. 2021, doi: 10.1039/D0EE02004D.

[80] J. Timo Wätjen, J. J. Scragg, M. Edoff, S. Rubino, and C. Platzer-Björkman, "Cu out-diffusion in kesterites - A transmission electron microscopy specimen preparation artifact," *Appl Phys Lett*, vol. 102, no. 5, Feb. 2013, doi: 10.1063/1.4790282/1068494.